\documentclass[preprint,3p, twocolumn]{elsarticle}
\usepackage[fleqn]{amsmath}
\usepackage{amssymb,amsthm}
\usepackage{mathtools}
\usepackage{mathrsfs}
\usepackage{bm}
\usepackage{slashed}	
\usepackage{cancel}
\usepackage{soul}
\usepackage{color}
\usepackage{xcolor}
\usepackage[pdftex,hypertexnames=false,linktocpage=true]{hyperref}
\hypersetup{colorlinks=true,linkcolor=red,anchorcolor=darkgreen,citecolor=green!50!blue,filecolor=darkgreen,urlcolor=green,
bookmarksnumbered=true,
pdfview=FitB
}
\usepackage{lineno}
\biboptions{sort&compress}
\usepackage{verbatim}
\usepackage{fullpage}
\usepackage{graphicx}
\usepackage{subfig} 	
\usepackage{relsize}	
\usepackage{float}
\usepackage{array}
\usepackage{booktabs}
\usepackage{arydshln}
\usepackage{multirow}

\makeatletter

\newcommand{\Rmnum}[1]{\expandafter\@slowromancap\romannumeral #1@}
\makeatother

\colorlet{darkgreen}{green!60!black}
\colorlet{brightyellow}{yellow!75!red}
\colorlet{orange}{red!50!yellow}
\colorlet{darkblue}{blue!60!black}
\colorlet{darkred}{red!80!black}
\colorlet{greenblue}{green!50!blue}

 \journal{Physics Letters B}

\begin{document}
\begin{frontmatter}

\title{Light mesons with one dynamical gluon on the light front}

\author[LU,imp,cas,impcas]{Jiangshan Lan}
\ead{jiangshanlan@impcas.ac.cn}
\author[imp,cas,impcas]{Kaiyu Fu}
\ead{kaiyufu94@gmail.com}
\author[imp,cas,impcas]{Chandan Mondal\corref{c1}}
\ead{mondal@impcas.ac.cn}
\author[imp,cas,impcas]{Xingbo Zhao}  
\ead{xbzhao@impcas.ac.cn}
\author[ISU]{James P. Vary}  
\ead{jvary@iastate.edu}

\author[]{\\\vspace{0.2cm}(BLFQ Collaboration)}
\cortext[c1]{Corresponding author}
\address[LU]{Lanzhou University, Lanzhou 730000, China}
\address[imp]{Institute for Modern Physics, Chinese Academy of Sciences, Lanzhou-730000, China} 
\address[cas]{School of Nuclear Science and Technology, University of Chinese Academy of Sciences, Beijing 100049, China}
\address[impcas]{CAS Key Laboratory of High Precision Nuclear Spectroscopy, Institute of Modern Physics, Chinese Academy of Sciences, Lanzhou 730000, China}

\address[ISU]{Department of Physics and Astronomy, Iowa State University, Ames, Iowa 50011, USA}

\begin{abstract}
We obtain the light meson mass spectroscopy from the light-front quantum chromodynamics (QCD) Hamiltonian, determined for their constituent quark-antiquark and quark-antiquark-gluon Fock components, together with a three-dimensional confinement. The eigenvectors of the light-front effective Hamiltonian provide a good quality description of the pion electromagnetic form factor, decay constant, and the valence quark distribution functions following QCD scale evolution. We also show that the pion's gluon densities can be probed through the pion-nucleus induced $J/\psi$ production data. Our pion parton distribution functions provide excellent agreement with $J/\psi$ production data from widely different experimental conditions.

\end{abstract}

\end{frontmatter}

\section{Introduction}
Quantum chromodynamics (QCD) is the well established theory for the strong interactions~\cite{Callan:1977gz}, where hadrons are viewed as confined systems of partons (quarks and gluons). However, understanding color confinement and chiral symmetry breaking remains incomplete and it is not yet possible to forecast, from QCD first principles, the extensive experimentally measured hadron spectroscopy. Successful theoretical frameworks for predicting some aspects of hadron spectra and  illuminating partonic structures are discretized space-time euclidean lattice~\cite{Joo:2019byq,Bazavov:2009bb,Fodor:2012gf,Briceno:2017max,Durr:2008zz,Hagler:2009ni} and the Dyson–Schwinger equations (DSEs) of QCD~\cite{Maris:2003vk,Roberts:1994dr,Alkofer:2000wg,Bashir:2012fs,Freese:2021zne}. Much progress is being made within the Hamiltonian formulation of QCD quantized on the light front (LF)~\cite{Brodsky:1997de,Bakker:2013cea}. Complementary insights into nonperturbative QCD can be achieved from LF holography~\cite{Brodsky:2014yha,Brodsky:2006uqa,deTeramond:2005su,deTeramond:2008ht,Branz:2010ub,Gutsche:2012ez,DeTeramond:2021jnn,Ahmady:2021lsh,Ahmady:2021yzh,Ahmady:2018muv,Lan:2020fno}. Meanwhile, basis light-front quantization (BLFQ) provides a computational framework for solving relativistic many-body bound state problems in quantum field theories~\cite{Vary:2009gt,Li:2021jqb,Zhao:2014xaa,Wiecki:2014ola,Li:2015zda,Jia:2018ary,Lan:2019vui,Lan:2019rba,Tang:2018myz,Tang:2019gvn,Xu:2019xhk,Xu:2021wwj,Lan:2019img,Qian:2020utg}.

With the framework of BLFQ~\cite{Vary:2009gt}, we consider an effective LF Hamiltonian and solve for its mass eigenvalues and  eigenstates at the scales suitable for low-resolution probes. With quarks and gluons being the explicit degrees of freedom for the strong interaction, our Hamiltonian incorporates LF QCD interactions~\cite{Brodsky:1997de} relevant to constituent quark-antiquark and quark-antiquark-gluon Fock components of the mesons with a complementary three-dimensional (3D) confinement~\cite{Li:2015zda}. By solving this Hamiltonian in the leading two Fock sectors and fitting the constituent parton masses and coupling constants as the model parameters, we obtain a good quality description of light meson mass spectroscopy. We evaluate the pion electromagnetic form factor (EMFF) and the  parton distribution functions (PDFs) from our resulting light-front wave functions (LFWFs) obtained as eigenvectors of this Hamiltonian. The former characterizes the spatial extent of a hadron, while the latter describe the longitudinal momentum distribution of partons within the hadron. Since we interpret our model as appropriate to a low energy scale, we apply QCD evolution of the PDFs to higher momentum scales, and compare results with experiment.

One salient issue can be addressed with our approach-the important role of an initial gluon distribution at a low energy scale that contributes to all (sea and gluon) the parton distributions under QCD scale evolution. Experimentally, the pion PDFs are measured with the pion-nucleus-induced Drell-Yan process~\cite{Conway:1989fs}, where the differential cross section is
sensitive to the valence quark densities at the parton's light-front momentum fraction $x > 0.2$. Sea and gluon
distributions contribute over a wider range of $x$. Meanwhile, pion-induced charmonium production is dominated by the quark-antiquark and gluon-gluon fusion partonic
processes so that the available $J/\psi$ production data~\cite{Gribushin:1995rt,Antoniazzi:1992af,Badier:1983dg,McEwen:1982fe} are sensitive to both the quark and gluon distributions of the incident pion~\cite{Chang:2020rdy}. Within the framework of the color evaporation model (CEM)~\cite{Einhorn:1975ua,Fritzsch:1977ay,Halzen:1977rs}, we study the sensitivity of $J/\psi$ production to our pion PDFs and confirm that consistency
with the data depends sensitively on the shape and magnitude of the pion's gluon density. Thus, the pion's initial gluon distribution at the valence quark regime can contribute significantly to interpreting the $J/\psi$ production data.

\section{Mass spectra and LFWFs}
The bound state problem on the LF is cast into an eigenvalue problem of the Hamiltonian: $P^-P^+|{\Psi}\rangle=M^2|{\Psi}\rangle$, where $P^\pm=P^0 \pm P^3$ employs the LF Hamitonian ($P^-$) and the longitudinal momentum ($P^+$) of the system, respectively, with the mass squared eigenvalue $M^2$. At fixed LF time ($x^+=t+z$), the meson state can be expressed in terms of various quark ($q$), antiquark ($\bar q$) and gluon $(g)$ Fock components,
\begin{align}\label{Eq1}
|\Psi\rangle=\psi_{(q\bar{q})}|q\bar{q}\rangle+\psi_{(q\bar{q}g)}|q\bar{q}g\rangle+\dots\, , 
\end{align}
where the LFWFs $\psi_{(\dots)}$ correspond to the probability amplitudes  to find different parton configurations in the meson. At the initial scale where the mesons are described by $|q\bar{q}\rangle$ and $|q\bar{q}g\rangle$, we adopt the LF Hamiltonian $P^-= P^-_{\rm QCD} +P^-_C$, where $P^-_{\rm QCD}$ and $P^-_{C}$ are the LF QCD Hamiltonian and a model for the confining interaction.  With one dynamical gluon in LF gauge~\cite{Brodsky:1997de}
\begin{align}
&P_{\rm QCD}^-= \int d^2 x^{\perp}dx^- \Big\{\frac{1}{2}\bar{\psi}\gamma^+\frac{m_{0}^2+(i\partial^\perp)^2}{i\partial^+}\psi\nonumber\\
& -\frac{1}{2}A_a^i\left[m_g^2+(i\partial^\perp)^2\right] A^i_a +g_s\bar{\psi}\gamma_{\mu}T^aA_a^{\mu}\psi \nonumber\\
&+\frac{1}{2}g_s^2\bar{\psi}\gamma^+T^a\psi\frac{1}{(i\partial^+)^2}\bar{\psi}\gamma^+T^a\psi \Big\},\label{eqn:PQCD}
\end{align}
where $\psi$ and $A^\mu$ are the quark and gluon fields, respectively. $T$ represents eight adjoint matrices of the $SU(3)$ gauge group, and $\gamma^+ = \gamma^0 + \gamma^3$, where $\gamma^\mu$ are the Dirac matrices.  The first two terms in Eq.~(\ref{eqn:PQCD}) are the kinetic energies of the quark and gluon, where $m_0$  and $m_g$ are the masses of the bare quark and gluon. While the gluon mass is zero in QCD, we allow a phenomenological gluon mass to be fit to the spectra in our low-energy model. The last two terms describe their interactions with coupling $g_s$. Following a renormalization procedure developed for positronium in a basis consisting of the $|e\bar{e}\rangle$ and $|e\bar{e}\gamma\rangle$~\cite{Zhao:2014hpa,Zhao:2020kuf}, we introduce a mass counter term, $\delta m_{q}= m_0 -m_{q}$ that represents the quark mass correction due to the quantum fluctuations to the higher Fock sector, where $m_{q}$ is the renormalized quark mass. Referring to Ref.~\cite{Glazek:1992aq}, we allow an independent quark mass $m_f$ in the vertex interaction. 

We adopt confinement in the leading Fock sector following~\cite{Li:2015zda}
\begin{align}
&P_{\rm C}^-P^+\nonumber\\
&=\kappa^4\left\{x(1-x) \vec{r}_{\perp}^2-\frac{\partial_{x}[x(1-x)\partial_{x}]}{(m_q+m_{\bar{q}})^2}\right\} \label{eqn:PC}
\end{align}
with $\kappa$ being the strength of the confinement. The transverse confinement corresponds to the LF holographic potential, where the holographic variable is $\sqrt{x(1-x)}\vec{r}_{\perp}$~\cite{Brodsky:2014yha}. The confining potential, Eq.~(\ref{eqn:PC}), reproduces a symmetric 3D confinement in the nonrelativistic limit and it has been successfully applied to mesons~\cite{Li:2021jqb,DeTeramond:2021jnn,Li:2015zda,Jia:2018ary,Lan:2019vui,Lan:2019rba,Tang:2018myz,Tang:2019gvn,Lan:2019img,Qian:2020utg} and to the nucleon~\cite{Xu:2019xhk}. Confinement in the $q\bar{q}g$ sector relies solely on the cutoffs of the BLFQ basis functions introduced below.
 
Following BLFQ, for each Fock-particle we employ a plane-wave, $e^{-i{ p^+ x^-/2}}$, to describe its longitudinal motion and two dimensional harmonic oscillator (``2D-HO") wave function, $\Phi_{nm}(\vec{p}_\perp;b)$ with scale parameter $b$, to describe its transverse degrees of freedom~\cite{Zhao:2014xaa}. The longitudinal motion is confined to a box of length $2L$ with antiperiodic (periodic)  boundary conditions for fermions (bosons). Therefore, the longitudinal momentum $p^{+}=2\pi k/L$, where $k=\frac{1}{2},\frac{3}{2},\frac{5}{2},...$ for fermions and $k=1,2,3,...$ for bosons. We neglect the zero mode for bosons. For all many-body basis states, we rescale the total longitudinal momentum $P^+=\sum_{i} p_{i}^{+}$ using $K=\sum_i k_i$ such that $P^+=\frac{2\pi}{L}K$.  For given parton $i$, the  longitudinal momentum fraction $x$ is then defined as $x_i=p^+_i/P^+=k_i/K$. The 2D-HO wave function carries the radial quantum number $n$ and angular quantum number $m$. With the additional quantum number $\lambda$ for the helicity, each single-parton basis state is identified using $\bar{\alpha}=\{x,n,m,\lambda\}$. In the case of Fock sectors allowing for multiple color-singlet state (i.e. beyond Fock sectors retained here) we need an additional label to distinguish each color singlet state. Additionally, our many-body basis states have well defined total angular momentum projection
$M_J=\sum_i\left(m_i+\lambda_i\right).$
We truncate the infinite basis space by introducing limits $K$ and $N_{\rm max}$, such that $\sum_i (2n_i+|m_i|+1)\le N_{\rm max}$ in longitudinal and transverse directions, respectively. The $N_{\rm max}$ truncation enables factorization of transverse center of mass motion~\cite{Wiecki:2014ola} and manifests a natural ultra-violet (UV) regulator $\Lambda_{\rm UV}\sim b\sqrt{N_{\rm max}}$.

The LFWFs of the mesons in momentum space are then expressed as
\begin{align}
&\Psi^{\mathcal{N},\,M_J}_{\{x_i,\vec{p}_{\perp i},\lambda_i\}}\nonumber\\
& =\sum_{ \{n_i m_i\} }\psi^{\mathcal{N}}({\{\overline{\alpha}_i}\})\prod_{i=1}^{\mathcal{N}}  \phi_{n_i m_i}(\vec{p}_{\perp i},b)\,,
\label{eqn:wf}
\end{align}
where $\psi^{\mathcal{N}=2}(\{\overline{\alpha}_i\})$ and $\psi^{\mathcal{N}=3}(\{\overline{\alpha}_i\})$ are the components of the eigenvectors relevant to the Fock sectors $|q\bar{q}\rangle$ and $|q\bar{q}g\rangle$, respectively, in the BLFQ basis obtained from diagonalizing the full Hamiltonian matrix. Meanwhile, $\phi_{n m}(\vec{p}_{\perp },b)$ represents the 2D-HO basis functions.

\begin{figure}
\begin{center}
\includegraphics[width=\linewidth]{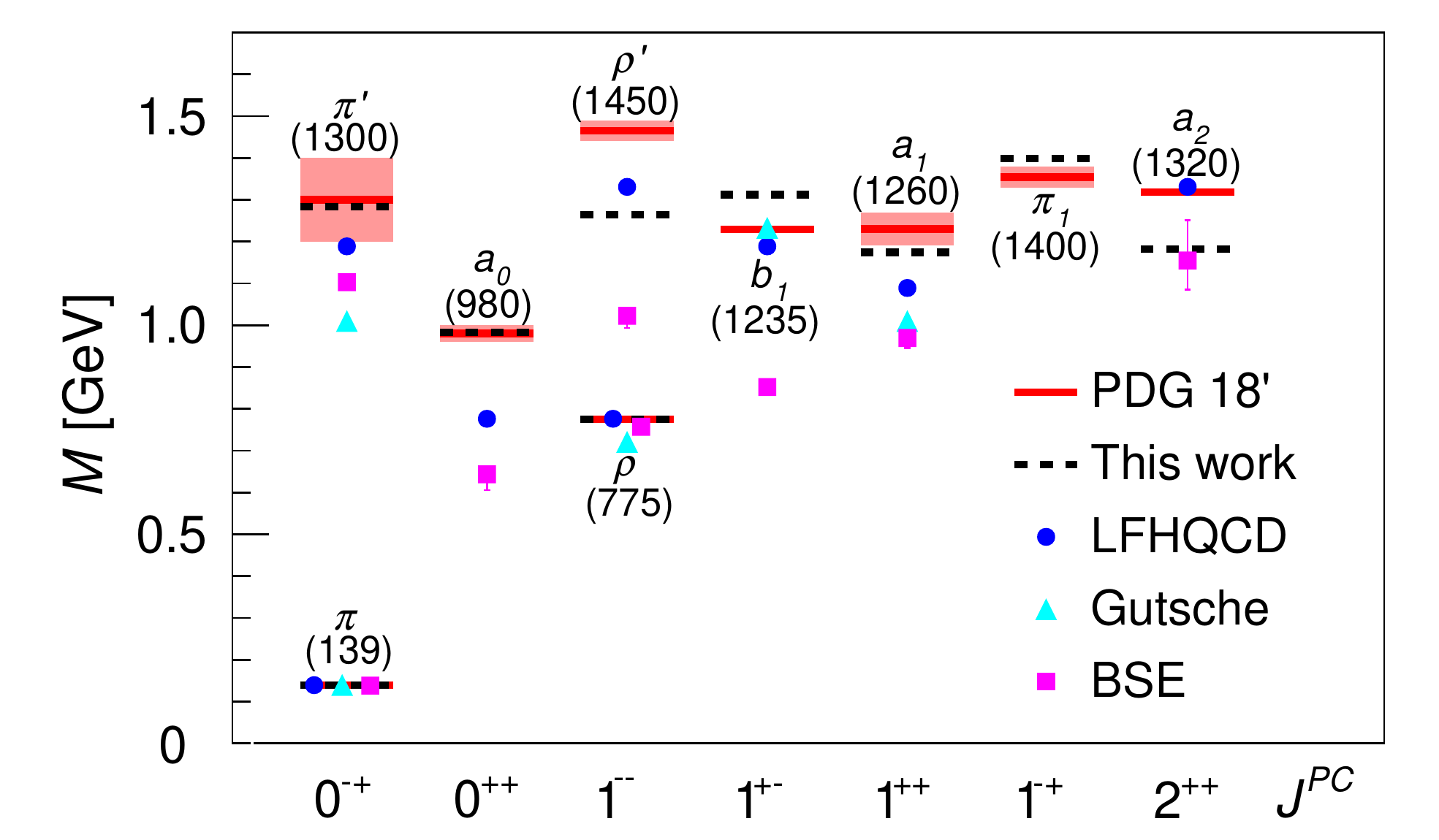}
\caption{The mass spectra of unflavored light mesons. Our results (Black-dashed bars) are compared with the experimental data taken from PDG (red-solid bars)  \cite{Tanabashi:2018oca}, the LFHQCD (blue-circle)~\cite{Brodsky:2014yha}, Gutsche {\it et. al.} (cyan-triangle)~\cite{Gutsche:2012ez}, and the BSE (magenta-square)~\cite{Fischer:2014xha}.}
\label{fig_mass}
\end{center}
\end{figure}

We select $\{N_{\rm max},K\}\,$=$\,\{14,15\}$, the HO scale parameter $b\,$=$\,0.29$ GeV and set our model parameters $\{m_q,\, m_g,\, \kappa,\, m_f,\, g_s  \}\,=\, \{{0.39 \,\rm GeV},\, {0.60 \,\rm GeV},\, {0.65\, \rm GeV}$, $\, {5.69\, \rm GeV},\, 1.92\}$ to fit the known masses of $\pi$, $\rho$, $a_0$, $a_1$, $\pi'$, and $\pi_1$. Figure~\ref{fig_mass} shows the mass spectra for unflavored light mesons and compares with experimental data compiled by the particle data group (PDG)~\cite{Tanabashi:2018oca} and with the predictions of LF holographic QCD (LFHQCD)~\cite{Brodsky:2014yha}. We also include results based on light-front holography~\cite{Gutsche:2012ez} and the Bethe Salpeter equation (BSE)~\cite{Fischer:2014xha} for comparison. For states not involved in the fit, $b_1$, $a_2$, and $\rho'$, our results deviate more significantly from the experimental data. However, we especially note that we are able to fit the hybrid state $\pi_1$, which may be viewed as a $q\bar{q}$ meson with a vibrating gluon flux tube~\cite{Tanabashi:2018oca}. 

\section{Pion EMFF and decay constant}
We first employ our resulting LFWFs for the pion EMFF $F(Q^2)$ via
\begin{align}
	\langle  \Psi (p^{\prime})| J_{\text{EM}}^\mu (0) | \Psi (p) \rangle = (p + p^{\prime})^\mu F(Q^2)\,,
\end{align}
where $p^{\prime}=p+q$, $Q^2=-q^2$ and the electromagnetic current $J_{\mathrm{EM}}^\mu(z)=\sum_f e_f\, \bar{q} (z) \gamma^\mu q(z)$ with $f=\bar{d},u$ and the quark electric charge $e_{f}$. Taking $\mu=+$, the EMFF can be expressed in terms of the meson LFWFs using the Drell-Yan-West formula \cite{Brodsky:2007hb},
\begin{align}
F(Q^2)=&\sum_f e_f \sum_{\mathcal{N},\,\lambda_i}\nonumber\\
 &\int_{\mathcal{N}}  \Psi^{\mathcal{N},\,M_J=0\,*}_{\{x_i,\vec{p}^{\,\prime}_{\perp i},\lambda_i\}}\,\Psi^{\mathcal{N},\,M_J=0}_{\{x_i,\vec{p}_{\perp i},\lambda_i\}}\,,
\label{eqn:ff_q}
\end{align}
where $\int_{\mathcal{N}} \equiv  \prod_{i=1}^\mathcal{N} \int \left[\frac{dx\,d^2\vec{p}_\perp}{16\pi^3}\right]_i 16\pi^3 \delta(1-\sum x_j)$ $\delta^2(\sum \vec{p}_{\perp j})$ and
for a struck parton, $\vec{p}^{\,\prime}_{\perp i}=\vec{p}_{\perp i}+ (1-x_i)\vec{q}_\perp$ while $\vec{p}^{\,\prime}_{\perp i}=\vec{p}_{\perp i}- x_i \vec{q}_\perp$ for the spectators. Considering the frame, where $q^+=0$, $Q^2=-q^2=\vec{q}^{\,2}_{\perp}$. The electric charge $e_u(e_{\bar{d}})=\frac{2}{3}(\frac{1}{3})$, while $e_g=0$. 
\begin{figure}
\begin{center}
\includegraphics[width=\linewidth]{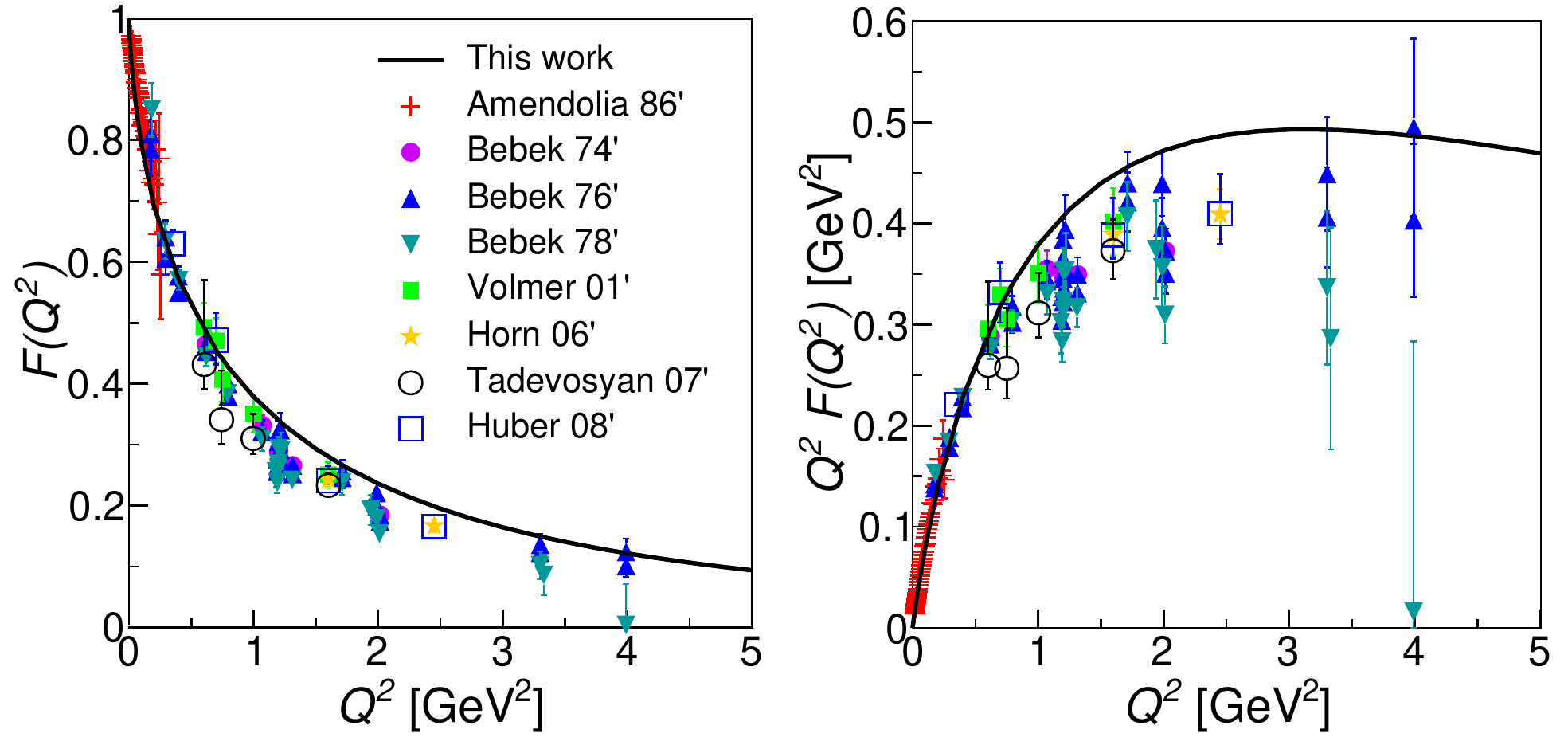}
\caption{The EMFF of the pion. The data are taken from Refs. \cite{Amendolia:1986wj,Bebek:1974iz,Bebek:1974ww,Bebek:1977pe,Volmer:2000ek,Horn:2006tm,Tadevosyan:2007yd,Huber:2008id}.
}
\label{fig_ff}
\end{center}
\end{figure}

Our prediction for the EMFF of the charged pion is compared to the experimental data in Fig.~\ref{fig_ff}. Note that our choice of $N_{\rm max}=14$, implies the UV regulator $\Lambda_{\rm UV}\sim b\sqrt{N_{\rm max}}\approx 1$ GeV. Our agreement with the precise low $Q^2$ EMFF data is consistent with our expectation that our predictions are most reliable in the low $Q^2$ regime.

The decay constant $f_{\rm P}$ of a pseudoscalar meson is defined as the local vacuum-to-hadron matrix element:
\begin{align}
	\langle  0|\, \bar{q}(0)\gamma^\mu\gamma_5 q(0)\,| \Psi (p) \rangle = ip^\mu f_{\rm P}\,.\label{decay}
\end{align}
We obtain $f_{\rm \pi}^{\rm Th.}=138$ MeV for the pion from the $|q\bar{q}\rangle$ Fock sector alone (the $|q\bar{q}g\rangle$ sector does not contribute) close to the experimental $f_{\rm \pi}^{\rm Exp.}=130.0\pm 0.2$ MeV~\cite{Tanabashi:2018oca}. 

{\it Pion PDFs.}---The PDF is the probability of finding a collinear parton carrying momentum fraction $x$. With our LFWFs, the valence quark (antiquark) and the gluon PDFs in the pion are given by
\begin{align}
&f_i(x)=\sum_{\mathcal{N},\,\lambda_i}\nonumber\\
& \int_{\mathcal{N}}  \Psi^{\mathcal{N},\,M_J=0\,*}_{\{x_i,\vec{p}_{\perp i},\lambda_i\}}\,\Psi^{\mathcal{N},\,M_J=0}_{\{x_i,\vec{p}_{\perp i},\lambda_i\}}\,\delta(x-x_i)\,,
\label{eqn:pdf_i}
\end{align}
where $i=q,\bar{q},g$ labels the valence quark, valence antiquark, and gluon, respectively. At our model scale the PDFs for the valence quark (antiquark) are normalized as $\int_0^1 f_{q/\bar{q}} (x)dx=1$, and those PDFs together with the gluon PDF satisfy the momentum sum rule: 
$\int_0^1 \sum_i x f_i(x) dx=1$.

We solve the next-to-next-to-leading order~(NNLO) Dokshitzer-Gribov-Lipatov-Altarelli-Parisi (DGLAP) equations \cite{Dokshitzer:1977sg,Gribov:1972ri,Altarelli:1977zs} of QCD numerically using the higher order perturbative parton evolution toolkit~\cite{Salam:2008qg} to evolve our PDFs from our model scale ($\mu_0^2$) to a higher scale ($\mu^2$).
We determine $\mu_0^2$ by requiring the result after evolution to produce the total first moments of the valence quark and the valence antiquark distributions from the global QCD analysis, $\langle x \rangle_{\rm valence}=0.48\pm0.01$ at $\mu^2=5$ GeV$^2$~\cite{Barry:2018ort}. This results in $\mu_0^2=0.34\pm 0.03$ GeV$^2$ and we then evolve our initial PDFs to the relevant experimental scale $\mu^2= 16~ {\rm GeV}^2$.  
While employing the DGLAP equations, we impose the condition
that the running coupling $\alpha_s({\mu^2})$ saturates in the infrared at
a cutoff value of max$\{\alpha_s\}\sim 1$~\cite{Lan:2020fno,Lan:2019vui,Lan:2019rba,Xu:2019xhk,Xu:2021wwj}. Note that the sea quark distributions are absent in the initial scales of our model.

\begin{figure}
\begin{center}
\includegraphics[width=0.95\linewidth]{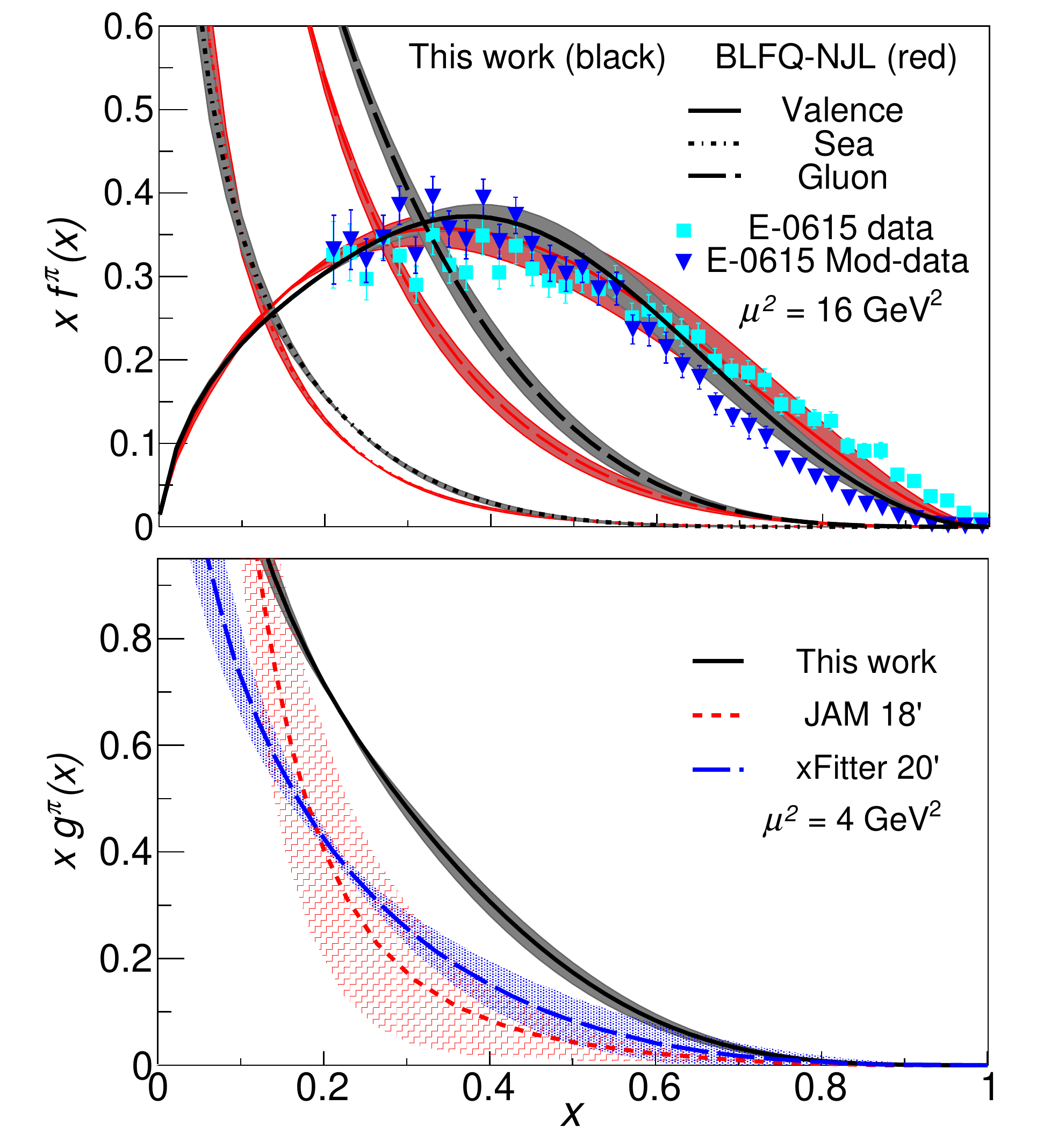}
\caption{The PDFs of the pion. Upper panel: the black lines are our results evolved from the initial scale $(0.34\pm 0.03~\mathrm{GeV}^2)$ using the NNLO DGLAP equations to the experimental scale of $16~\mathrm{GeV}^2$. 
The red lines correspond to BLFQ-NJL predictions~\cite{Lan:2019vui}. Results are compared with the original analysis of the FNAL-E615 experiment~\cite{Conway:1989fs} data and with its reanalysis (E615 Mod-data)~\cite{Chen:2016sno}. Lower panel: Our result for the pion gluon PDF at $\mu^2=4$ GeV$^2$ is compared with the global fits, JAM~\cite{Barry:2018ort} and xFitter~\cite{Novikov:2020snp}.}
\label{fig_pdf}
\end{center}
\end{figure}

Figure~\ref{fig_pdf} shows our results for the pion PDFs, where we compare the valence quark distribution after QCD evolution with the data from the E615 experiment~\cite{Conway:1989fs} as well as the reanalysis of the E615 experiment~\cite{Chen:2016sno,Aicher:2010cb}. We also include the pion PDFs previously obtained in BLFQ-NJL model~\cite{Jia:2018ary,Lan:2019vui,Lan:2019rba} based on a valence Fock representation for comparison. The error bands in our evolved PDFs are manifested from an adopted $10\%$ uncertainty in our initial scale. We find a good agreement between our prediction for the pion valence quark distribution and the  
reanalyzed E615 data, while the BLFQ-NJL model prefers the original E615 data. In our current treatment, the pion valence PDF falls off as $(1-x)^{1.77}$, favoring a slightly slower falloff compared to the $(1-x)^2$ predicted by perturbative QCD~\cite{Berger:1979du} and calculations using DSEs~\cite{Hecht:2000xa}. Note that Refs.~\cite{Chen:2016sno,Aicher:2010cb,Bednar:2018mtf} support our finding. On the other hand, the BLFQ-NJL at same high-$x$ exhibits $(1-x)^{1.44}$~\cite{Lan:2019vui}, supported by Ref.~\cite{deTeramond:2018ecg}.

The gluon distribution significantly increases in our approach compared to that in the BLFQ-NJL model as well as to the global fits~\cite{Barry:2018ort,Novikov:2020snp} as can be seen from Fig.~\ref{fig_pdf}. The BLFQ-NJL model is based on the pion valence Fock sector and gluons are generated solely from the scale evolution. However, our model includes a dynamical gluon at the initial scale and results in a larger gluon PDF at large-$x~ (>0.2)$ after scale evolution. Our result is also supported by a recent study from DSEs~\cite{Freese:2021zne}. 
We find that gluons carry $\{39.5,\,42.1,\,43.9,\,44.6,\,45.1\}\%$ of pion momentum at $\{1.69,\,4,\,10,\,16,\,27\}$ GeV$^2$, respectively.

\section{Pion-nucleus induced $J/\psi$ production}
Finally, we perform the next-to-leading order (NLO) calculations of the differential cross sections for charmonium production by a pion beam to compare with available $J/\psi$ production data with a special focus on the role of our model's gluon distribution. Among several theoretical approaches~\cite{Einhorn:1975ua,Fritzsch:1977ay,Halzen:1977rs,Chang:1979nn,Berger:1980ni,Baier:1983va,Bodwin:1994jh}, we adopt the CEM~\cite{Einhorn:1975ua,Fritzsch:1977ay,Halzen:1977rs}, assuming a constant probability for $c\bar{c}$ pairs to hadronize into a given charmonium. The CEM has only one effective parameter ($F$) that accounts for the probability and provides a good description of many features of fixed-target $J/\psi$ cross section data with proton beams~\cite{Gavai:1994in,Schuler:1996ku} and the collider data at the Relativistic Heavy Ion Collider and Large Hadron Collider~\cite{Nelson:2012bc,Lansberg:2020rft}.

The differential cross section for $J/\psi$ production from the pion-nucleus collision in the CEM is given by \cite{Nason:1987xz,Nason:1989zy,Mangano:1992kq,Chang:2020rdy}
\begin{align}
&\frac{d{\bf \sigma}}{dx_{\mathrm{F}}}\Big\vert_{J/\psi}\\
 &=F\sum_{i,j=q,\bar{q},g}\int^{2m_D}_{2m_c}dM_{c\bar{c}}\frac{2M_{c\bar{c}}}{S\sqrt{x_F^2+4M^2_{c\bar{c}}/S}}\nonumber\\
&\times \hat{\bf \sigma}_{ij}(s,m^2_{c},\mu_F^2,\mu_R^2)\,f^{\pi^\pm}_i(x_1,\mu_F^2)\,{f}^N_{j}(x_2,\mu^2_F)\,\nonumber\label{crosseq}
\end{align}
with $x_{1,2}=(\sqrt{x_F^2+4M^2_{c\bar{c}}/S}\pm x_F)/2$ and $x_F=x_1-x_2$. The variables $S$ and $s$ denote the square of center-of-mass energy of the colliding $\pi N$ and the interacting partons, respectively, and  
$m_c$, $m_D$, and $M_{c\bar{c}}$ are the masses of the charm quark, $D$ meson, and $c\bar{c}$ pair, respectively. The $\hat{\sigma}_{ij}$ is the short-distance differential cross section of heavy quark pair production at NLO, calculable as a perturbation series in the strong coupling \cite{Nason:1987xz}. The $f^{\pi^\pm}$ and $f^{N}$ are the PDFs of the pion and the target nuclei, respectively, evaluated at the factorization scale $\mu_F$. To complement our model for the pion PDFs, we adopt the nuclear PDFs from nCTEQ 2015~\cite{Kovarik:2015cma} and consider the factorization scale $\mu_F=2m_c$ and the renormalization scale $\mu_R=m_c$ with $m_c= 1.5$ GeV to perform the calculation for the cross section~\cite{Mangano:1992kq}.

\begin{figure}
	\begin{center}
			\includegraphics[width=\linewidth]{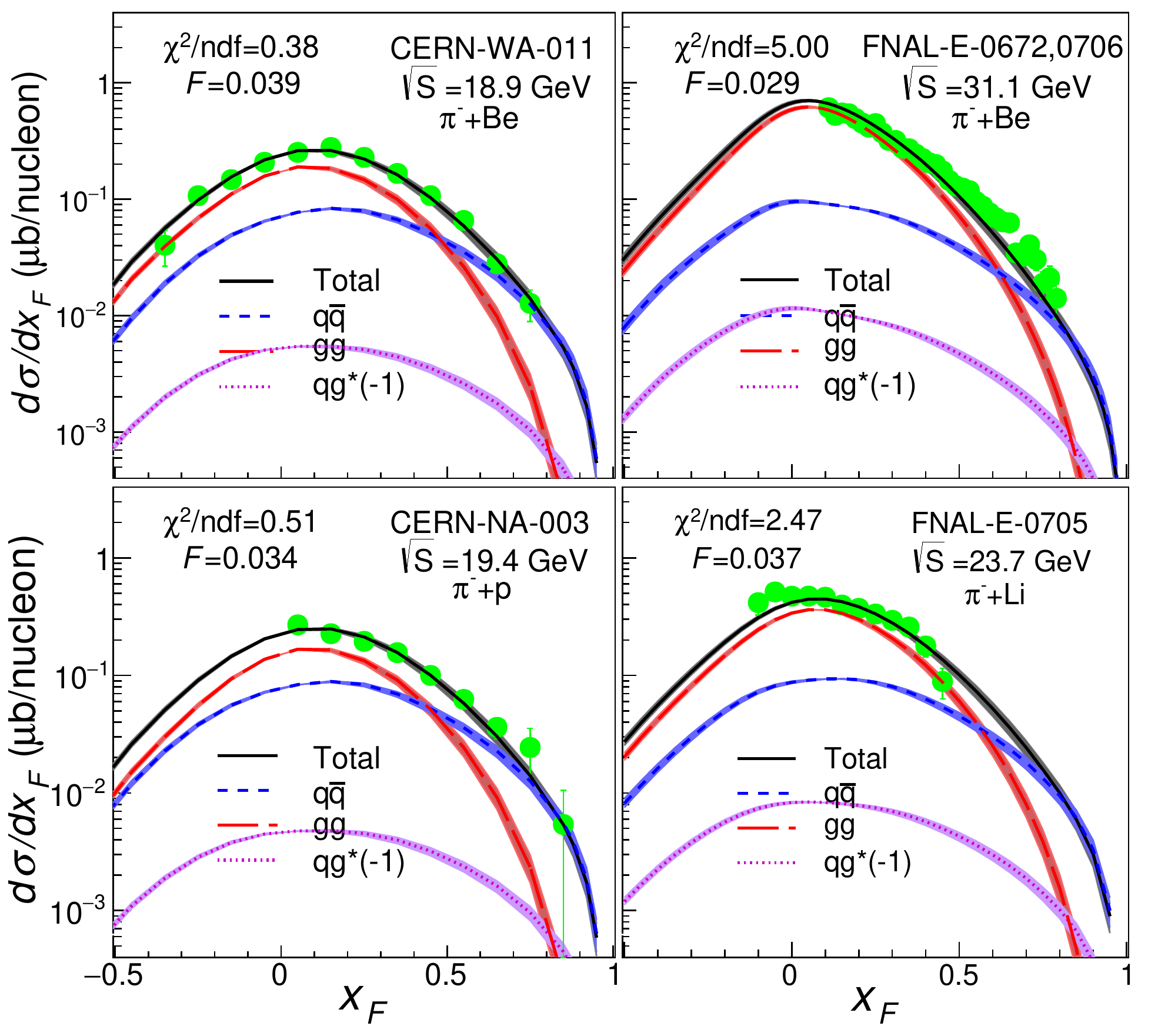}
		\caption{$ d{\bf \sigma}/dx_{F}$ for the ${\pi^-\rm{nucleus}}\rightarrow{J/\psi X}$ process as a function of $x_F$. The data are taken from FNAL E672, E706, E705,  CERN NA003 and  CERN WA11 experiments~\cite{Gribushin:1995rt,Antoniazzi:1992af,Badier:1983dg,McEwen:1982fe}. The negative contribution from $qg$ is presented here by multiplying with ``$(-1)$".}
		\label{fig_cross}
	\end{center}
\end{figure}

Figure~\ref{fig_cross} illustrates the cross section $d\sigma/dx_F$ for the $J/\psi$ production as a function of $x_F$ and compares with various experimental data, including the E672 and E706 experiments with 515 GeV pions and beryllium target \cite{Gribushin:1995rt}, the E705 experiment with 300 GeV pions and lithium target \cite{Antoniazzi:1992af}, the NA3 experiment with 200 GeV pions and hydrogen target \cite{Badier:1983dg}, and the WA11 experiment with 190 GeV pions and beryllium target \cite{McEwen:1982fe}. We determine the hadronization factor $F$, as an overall normalization parameter, by the best $\chi^2$ fit to those data, shown as the black lines in Fig.~\ref{fig_cross}. The values of $F$ and the corresponding $\chi^2/$ndf values of all the best fits are also displayed in the plot (``ndf" represents ``number of degrees-of-freedom"). We obtain good $\chi^2/$ndf values for all four data sets. We also present the individual $q\bar{q}$, $gg$ and $qg$ contributions to the total cross sections in the NLO calculation. Notably, the $gg$ contribution dominates the cross section up to $x_F\sim 0.5$ and it decreases dramatically toward larger $x_F$. The contribution from $gg$ is increasing as the energy increases. Meanwhile, the $q\bar{q}$ contribution exhibits slower falloff and the relative importance of $q\bar{q}$ rises at the large-$x_F$ region. This is because the valence quark (antiquark) dominates the PDFs at large-$x$. The contribution from $qg$ is negative (as signified by the ``$(-1)$" in the labels) and relatively small. Overall, we observe that the $J/\psi$ production data are sensitive to the pion PDFs, especially the large-$x$ ($x>0.2$) gluon distribution of pions. Our pion PDFs provide good agreement with data from widely different experimental conditions. Our finding also supports the study reported in Ref.~\cite{Chang:2020rdy}.

\section{Conclusion and outlook}
We have solved the light-front QCD Hamiltonian for the light mesons by considering them within the constituent quark-antiquark and the quark-antiquark-gluon Fock spaces. Together with a three-dimensional confinement in the leading Fock sector, the eigenvalues of the Hamiltonian in Basis Light Front Quantization provide a good quality description of the light mesons' mass spectra. The LFWFs obtained as the eigenvectors of this Hamiltonian were then employed to produce the pion EMFF and the initial PDFs. We have obtained excellent agreement with the experimental data in the low-$Q^2$ regime for the pion EMFF. The PDFs at a higher experimental scale have been computed based on the NNLO DGLAP equations and we obtain reasonable agreement with the experimental data for the valence quark distribution. 

We have also calculated the differential cross sections for pion-induced $J/\psi$ production using the CEM framework at NLO and have obtained good agreement with available data~\cite{Gribushin:1995rt,Antoniazzi:1992af,Badier:1983dg,McEwen:1982fe}. Our study indicates that a proper description of $J/\psi$ production data imposes a strong constraint on the pion's PDFs. 
We confirm the sensitivity of the $J/\psi$ production data to the pion's gluon density in the valence-quark regime. The resulting LFWFs can be employed to study other quark and gluon distributions, such as the generalized parton distributions,
the transverse momentum dependent parton distributions as well as the double parton correlations etc., in the light mesons. On the other hand, this work can be extended to higher Fock sectors to incorporate, for example, sea degrees of freedom as well.

\section*{Acknowledgements}
We thank J. Wu, H. Zhao, S. Xu, S. Jia and S. Platchkov for useful discussions. C. M. thanks the Chinese Academy of Sciences President's International Fellowship Initiative for the support via Grants No. 2021PM0023. C. M. is supported by new faculty start up funding by the Institute of Modern Physics, Chinese Academy of Sciences, Grant No. E129952YR0. X. Z. is supported by new faculty startup funding by the Institute of Modern Physics, Chinese Academy of Sciences, by Key Research Program of Frontier Sciences, Chinese Academy of Sciences, Grant No. ZDB-SLY-7020, by the Natural Science Foundation of Gansu Province, China, Grant No. 20JR10RA067 and by the Strategic Priority Research Program of the Chinese Academy of Sciences, Grant No. XDB34000000. J. P. V. is supported by the Department of Energy under Grants No. DE-FG02-87ER40371, and No. DE-SC0018223 (SciDAC4/NUCLEI). A portion of the computational resources were also
provided by Gansu Computing Center.


\end{document}